\documentstyle[12pt,fleqn]{article}
\def\beeq{\begin{equation}}
\def\eneq{\end{equation}}
\def\beeqa{\begin{eqnarray}}
\def\eneqa{\end{eqnarray}}

\addtocounter{section}{-1}
\begin{document}
\begin{center}

\mbox{}

\vspace{1cm}

{\large {\bf {
Coulomb interaction effects on nonlinear optical response\\
in C$_{\bf 60}$, C$_{\bf 70}$, and higher fullerenes
} } }

\vspace{1cm}

Kikuo Harigaya\footnote[1]{E-mail address: 
\verb+harigaya@etl.go.jp+; URL: 
\verb+http://www.etl.go.jp/+\~{}\verb+harigaya/+}

\vspace{1cm}

{\sl Electrotechnical Laboratory, Tsukuba 305, Japan}

\end{center}
\vspace{1cm}

\noindent
{\bf Abstract}\\
Nonlinear optical properties in the fullerene C$_{60}$ and 
the extracted higher fullerenes -- C$_{70}$, C$_{76}$, C$_{78}$, 
and C$_{84}$ -- are theoretically investigated by using the
exciton formalism and the sum-over-states method.  We find 
that off-resonant third order susceptibilities of higher 
fullerenes are a few times larger than those of C$_{60}$.  
The magnitude of nonlinearity increases as the optical gap 
decreases in higher fullerenes.  The nonlinearity is nearly 
proportional to the fourth power of the carbon number when 
the onsite Coulomb repulsion is $2t$ or $4t$, $t$ being the 
nearest neighbor hopping integral.  This result, indicating 
important roles of Coulomb interactions, agrees with quantum 
chemical calculations of higher fullerenes.

\pagebreak

It has been found that C$_{60}$ thin films show large optical 
nonlinearities [1-4] which are attractive from the viewpoint of 
scientific interest as well as technological applications.  
The magnitude of the third-order nonlinear susceptibility, 
$\chi^{(3)}_{\rm THG} (\omega) = \chi^{(3)} 
(3\omega;\omega,\omega,\omega)$, for third-harmonic generation 
(THG) is of the order of $10^{-12}$ esu to $10^{-11}$ esu.  
This large response is comparable to the responses measured 
for polydiacetylenes.  The optical spectra of C$_{70}$ [4] and
higher fullerenes (C$_{76}$, C$_{78}$, C$_{84}$, etc.) [5,6] 
have been obtained.  In order to explain the results of several 
interesting experiments, theoretical investigations [7-11]
have been performed.  We have applied a tight binding model [7]
to C$_{60}$, and have analyzed the nonlinear optical properties 
of C$_{60}$.  Coulomb interaction effects on the absorption 
spectra and the optical nonlinearity of C$_{60}$ have also 
been studied [10].  We have found that the linear absorption 
spectra of C$_{60}$ and C$_{70}$ are well explained by the 
Frenkel exciton picture [11] except for the charge transfer 
exciton feature around the excitation energy of 2.8 eV of 
the C$_{60}$ solids [12].  Coulomb interaction effects reduce 
the magnitude of the optical nonlinearity of C$_{60}$ compared 
with that determined using the free electron calculation [10], 
and we have discussed the possibility that the local field 
enhancement might be effective in solids.

In this paper, we investigate nonlinear optical properties of 
higher fullerenes.   We focus on the off-resonant third order 
susceptibility in order to estimate the magnitudes of the 
nonlinear optical responses of each isomer.  The Coulomb 
interaction strengths are also changed in a reasonable range, 
because realistic strengths are not well known in higher 
fullerenes.  The Coulomb interactions are taken into account 
by the parametrized Ohno potential, 
$W(r) = 1 / \sqrt{ (1/U)^2 + (r/r_0 V)^2 }$, between two 
electrons with distance $r$.  Here, $U$ is the interaction 
strength at a distance $r=0$, $V$ is the strength of the 
long range part, and $r_0$ is the mean bond length.  
Based on our results for the optical properties of 
C$_{60}$ and C$_{70}$ [10,11], we can assume $V = U/2$.  
The onsite Coulomb strength is varied within the range 
$0 \leq U \leq 4t$, $t$ being the hopping integral between
nearest neighbor carbon atoms.

Figure 1 shows the absolute value of the off-resonant 
susceptibility, $\chi^{(3)} (0) = \chi^{(3)} (0;0,0,0)$, 
plotted against the Coulomb interaction strength $U$.  
The different plots indicate different types of isomers.  
The four isomers -- C$_{70}$, $D_2$-C$_{76}$, $D_3$-C$_{78}$, 
and one type of $C_{2v}$-C$_{78}$ isomer [$C_{2v}$ by Kikuchi 
et al's notation (Ref. 13)] -- exhibit similar magnitudes of 
optical nonlinearities which are larger than those of C$_{60}$.
On the other hand, the other three isomers -- another type of 
$C_{2v}$-C$_{78}$ isomer [$C_{2v}^{'}$ by Kikuchi et al's notation (Ref. 13)], 
$D_{2d}$-C$_{84}$, and $D_2$-C$_{84}$ -- show larger optical 
nonlinearities than those of the first four isomers.  This is 
mainly due to the smaller energy gap of the latter isomers, 
even though the negative correlation between the susceptibility 
and the energy gap is not so complete through all the isomers.
The decrease in the susceptibility between the free electron model 
($U = 0$) and the case in which $U=4t$ is by a factor of approximately 
0.1 for all the isomers, indicating that this is a general 
property of various kinds of higher fullerenes.  The overall 
magnitudes of the susceptibility are around 10$^{-12}$ esu for 
most of the Coulomb interactions considered.

In Fig. 2, the relations between the absolute value of the 
off-resonant susceptibility and the energy gap are shown for 
three Coulomb interaction strengths: $U = 0t$, $2t$, and $4t$.  
Here, the energy gap is defined as the optical excitation 
energy of the lowest dipole allowed state, in other words, 
the optical gap.  For each Coulomb interaction, the plots 
(squares, circles, or triangles) cluster in a bunch.  When the 
energy gap becomes larger, the susceptibility tends to decrease.  
However, the correlation between the susceptibility and the 
energy gap is far from that of a smooth function.  The 
correlation is merely a kind of tendency.  Therefore, the 
decrease in the energy gap of higher fullerenes is one origin 
of the larger optical nonlinearities of the systems.  The 
actual magnitudes of nonlinearities would also be influenced 
by the detailed electronic structures of isomers.

In the calculations for C$_{60}$ reported previously, the 
magnitudes of the THG at the energy zero are approximately 
$1 \times 10^{-12}$ esu in the free electron model [7], and 
approximately $2 \times 10^{-13}$ esu for $U = 4t$ and 
$V = 2t$ [10].  These results have been shown in 
Fig. 1, also.  In the present calculations for higher 
fullerenes, the magnitudes are a few times larger than those 
of C$_{60}$.  Thus, the author predicts that nonlinear 
optical responses in higher fullerenes are generally larger 
than in C$_{60}$.  In our previous paper [10], we discussed 
the fact that the local field correction factor is of the 
order of 10 for C$_{60}$ solids.  Since the distance between 
the surfaces of neighboring fullerene molecules in C$_{70}$ 
and C$_{76}$ solids is nearly the same as in C$_{60}$ solids, 
we expect that local field enhancement in thin films of higher 
fullerenes is of a magnitude similar to that in C$_{60}$ systems.

It is of some interests to look at carbon number dependence
of the magnitude of the optical nonlinearity of the calculated
isomers in higher fullerenes.  Figure 3 shows $|\chi^{(3)}(0)|$
as functions of the carbon number $N$ for three Coulomb interaction
strengths, $U=0t$, $2t$, and $4t$.  The solid lines indicate
the linear fitting in the logarithmic scale: $|\chi^{(3)}(0)|
\sim A \cdot N^\alpha$.  The powers $\alpha$ for the three Coulomb
interaction strengths are summarized in TABLE I.  When $U=0t$, 
the power $\alpha$ is about 5.  As $U$ increases, $\alpha$
decreases.  It is among 4, when $U \sim 2t$ and $4t$.  This
magnitude of the power 4 agrees with the result of the quantum
chemical calculation of higher fullerenes upto C$_{84}$ [14].
Therefore, we have shown important roles of Coulomb interactions
in nonlinear optical response of higher fullerenes.

Experimental measurements of optical nonlinearities in higher
fullerenes whose carbon number is larger than 70 have not been
reported so much, possibly because of the difficulty in obtaining
samples with good quality and the difficult measurements.  However,
the recent report of the degenerate four-wave-mixing measurement
of C$_{90}$ in solutions [15] indicates the larger optical nonlinearity
than that in C$_{60}$.  The magnitude of $\chi^{(3)}$ is about
eight times larger than in C$_{60}$, and is apparently enhanced
from that of the theoretical predictions: $(90/60)^4 = 1.5^4 = 5.063$.
Therefore, further experimental as well as theoretical investigations
of nonlinear optical properties in higher fullerenes should be
fascinating among scientists and technologists of the field of
photophysics.

In summary, we have investigated the nonlinear optical properties
of higher fullerenes.  Theoretical off-resonant third order
susceptibility has been calculated using the exciton theory.  
We have found optical nonlinearities of higher fullerenes which
are larger than those of C$_{60}$.  The magnitude of $\chi^{(3)}$ 
tends to increase as the optical gap decreases in higher fullerenes.
The nonlinearity is nearly proportional to the fourth power of 
the carbon number when the onsite Coulomb repulsion is $2t$ or 
$4t$.  This result, indicating important roles of Coulomb 
interactions, agrees with quantum chemical calculations of 
higher fullerenes.

\pagebreak

\noindent
{\bf References}

\mbox{}

\noindent
$[1]$ J. S. Meth, H. Vanherzeele and Y. Wang, Chem. Phys. Lett. 
{\bf 197} (1992) 26.\\
$[2]$ Z. H. Kafafi. J. R. Lindle, R. G. S. Pong, F. J. Bartoli, 
L. J. Lingg and J. Milliken, Chem. Phys. Lett. {\bf 188} (1992) 492.\\
$[3]$ F. Kajzar, C. Taliani, R. Danieli, S. Rossini and R. Zamboni,
Chem. Phys. Lett. {\bf 217} (1994) 418.\\
$[4]$ B. C. Hess, D. V. Bowersox, S. H. Mardirosian and
L. D. Unterberger, Chem. Phys. Lett. {\bf 248} (1996) 141.\\
$[5]$ R. Ettl, I. Chao, F. Diederich and R. L. Whetten, Nature 
{\bf 353} (1991) 149.\\
$[6]$ K. Kikuchi, N. Nakahara, T. Wakabayashi, M. Honda, H. Matsumiya,
T. Moriwaki, S. Suzuki, H. Shiromaru, K. Saito, K. Yamauchi,
I. Ikemoto and Y. Achiba, Chem. Phys. Lett. {\bf 188} (1992) 177.\\
$[7]$ K. Harigaya and S. Abe, Jpn. J. Appl. Phys. {\bf 31} (1992) L887.\\
$[8]$ E. Westin and A. Ros\'{e}n, Int. J. Mod. Phys. B {\bf 23-24}
(1992) 3893.\\
$[9]$ Z. Shuai and J. L. Br\'{e}das, Phys. Rev. B {\bf 46} (1992) 16135.\\
$[10]$ K. Harigaya and S. Abe, J. Lumin. {\bf 60\&61} (1994) 380.\\
$[11]$ K. Harigaya and S. Abe, Phys. Rev. B {\bf 49} (1994) 16746.\\
$[12]$ S. L. Ren, Y. Wang, A. M. Rao, E. McRae, J. M. Holden, 
T. Hager, K. A. Wang, W. T. Lee, H. F. Ni, J. Selegue 
and P. C. Eklund, Appl. Phys. Lett. {\bf 59} (1991) 2678.\\
$[13]$ K. Kikuchi, N. Nakahara, T. Wakabayashi, S. Suzuki,
H. Shiromaru, Y. Miyake, K. Saito, I. Ikemoto, M. Kainosho
and Y. Achiba, Nature {\bf 357} (1992) 142.\\
$[14]$ M. Fanti, G. Orlandi and F. Zerbetto,
J. Am. Chem. Soc. {\bf 117} (1995) 6101.\\
$[15]$ H. Huang, G. Gu, S. Yang, J. Fu, P. Yu, G. K. L. Wong 
and Y. Du, Chem. Phys. Lett. {\bf 272} (1997) 427.\\

\pagebreak

\noindent
TABLE I.  Coulomb interaction dependence of the power $\alpha$
where $|\chi^{(3)}(0)| \sim A \cdot N^\alpha$.

\mbox{}

\begin{tabular}{cc} \hline \hline
$U$  & $\alpha$ \\ \hline
0$t$ & 5.253 \\
2$t$ & 4.133 \\
4$t$ & 3.536 \\ \hline \hline
\end{tabular}

\pagebreak

\begin{flushleft}
{\bf Figure Captions}
\end{flushleft}

\mbox{}

\noindent
Fig. 1.  The absolute value of the off-resonant susceptibility
$| \chi^{(3)} (0)|$ plotted against the Coulomb interaction 
strength $U$.  The stars represent numerical results for C$_{60}$.
The closed and open squares represent results for C$_{70}$ and 
$D_2$-C$_{76}$, respectively.  The closed circles represent 
results for $D_3$-C$_{78}$, and the open and crossed circles 
represent results for two types of $C_{2v}$-C$_{78}$.  The 
closed and open triangles represent results for $D_{2d}$-C$_{84}$ 
and $D_2$-C$_{84}$, respectively.

\mbox{}

\noindent
Fig. 2.  The absolute value of the off-resonant susceptibility
$| \chi^{(3)} (0)|$ for C$_{60}$ and seven isomers of
higher fullerenes, plotted against the energy
gap (shown in units of $t$).  The squares, circles, and
triangles represent results for $U = 0t$, $2t$, and $4t$, 
respectively.  The left axis is in the logarithmic scale.

\mbox{}

\noindent
Fig. 3.  The absolute value of the off-resonant susceptibility
$| \chi^{(3)} (0)|$ for C$_{60}$ and seven isomers of
higher fullerenes, plotted against the carbon number $N$.  
The squares, circles, and triangles represent results for 
$U = 0t$, $2t$, and $4t$, respectively.  The left and bottom
axes are in the logarithmic scale.  The solid lines are
the results of the linear fitting in the logarithmic scale: 
$|\chi^{(3)}|\sim A \cdot N^\alpha$.

\end{document}